\documentclass[prb,twocolumn,amssymb,floatfix,amsmath]{revtex4}

\usepackage{amsfonts}
\usepackage{bm}
\usepackage[dvips]{color}
\usepackage{graphicx}
\usepackage{subfig}
\usepackage{epstopdf}

\newcommand{\bK}{{\bm K}}
\newcommand{\bKp}{{{\bm K}_+}}
\newcommand{\bKm}{{{\bm K}_-}}
\newcommand{\bb}{{\bm b}}
\newcommand{\bd}{{\bm \Delta}}

\newcommand{\br}{{\bm r}}
\newcommand{\ldos}{LDOS }
\newcommand{\vsig}{\vec{\sigma}}
\newcommand{\grad}{\vec{\nabla}}
\newcommand{\A}{\vec{A}}
\newcommand{\vef}{\vec{e_\varphi}}
\newcommand{\vLp}{\vec{L}'}
\newcommand{\veone}{\vec{e}_1}
\newcommand{\vetwo}{\vec{e}_2}

\begin{document}

\title{Electronic States of Graphene Grain Boundaries}

\author{A. Mesaros$^1$, S. Papanikolaou$^2$, C. F. J. Flipse$^3$, D. Sadri$^1$$\footnote{Present address: Institute of Theoretical Physics, Ecole Polytechnique F\'ed\'erale de Lausanne (EPFL), CH-1015 Lausanne, Switzerland}$ and J. Zaanen$^1$}

\affiliation{$^1$Instituut--Lorentz, Universiteit Leiden, P. O. Box 9506, 2300 R A Leiden, The Netherlands}
\affiliation{$^2$LASSP, Physics Department, Clark Hall, Cornell University, Ithaca, NY 14853-2501}
\affiliation{$^3$Department of Applied Physics, Eindhoven University of Technology, 5600 MB Eindhoven, The Netherlands}

\begin{abstract}
We introduce a model for amorphous grain boundaries in graphene, and find that stable structures can exist along the boundary that are responsible for local density of states enhancements both at zero and finite ($\sim 0.5$ eV) energies. Such zero energy peaks in particular were identified in STS measurements [J. \v{C}ervenka, M. I. Katsnelson, and C. F. J. Flipse, Nature Physics 5, 840 (2009)], but are not present in the simplest pentagon-heptagon dislocation array model [O. V. Yazyev and S. G. Louie, Physical Review B 81, 195420 (2010)]. We  consider the low energy continuum theory of arrays of dislocations in graphene and show that it predicts localized zero energy states. Since the continuum theory is based on an idealized lattice scale physics it is a priori not literally applicable. However, we identify stable dislocation cores, different from the pentagon-heptagon pairs, that do carry zero energy states. These might be responsible for the enhanced magnetism seen experimentally at graphite grain boundaries.
\end{abstract}

\date{\today}

\maketitle

\section{Introduction}

Grain boundaries and other extended defect structures in graphite have been studied by surface measurements techniques for quite some time.~\cite{Simonis:2002p1044,Pong:2007p834,Albrecht:1988p3433,Snyder:1992p919} This research actually reaches beyond the fundamental questions of mechanical material properties and crystalline ordering complexities. The study of defects on the surface layer of graphite are directly related to the influence of disorder on isolated graphene sheets, and thereby of direct relevance in the context of graphene's extraordinary properties and potential electronic applications. Grain boundaries 
have a special status, since they are the natural extended defects also in  two-dimensional graphene, while they have a topological status  since in terms of the lattice order they can be represented as an array of dislocations with Burgers vectors that do not cancel.~\cite{Yazyev:2010p3429}

The STS studies of graphite have also revealed some clues about the connection of extended defects and the controversial ferromagnetic properties of metal-free carbon.~\cite{Cervenka:2009p2555}  Earlier theoretical studies aiming at localized defects in graphene,~\cite{Vozmediano:2005p2562,Peres:2006p285} do yield some insights into the electronic states and magnetic properties of some types of graphene edges, cracks, and single atom defects. However, theoretical studies of the extended defect structures themselves have been completely absent until recently.~\cite{Yazyev:2010p3429}

The recent STM and STS studies of the electronic properties of defect arrays in graphite~\cite{Cervenka:2008p2547,Cervenka:2009p2555} have shown that the local density of states (\ldos) has two types of characteristic features: either an enhancement at zero energy, or a pair of peaks at low energy below symmetrically distributed around the Fermi energy. A first-principles model of grain boundaries based on a periodic array of the simplest pentagon-heptagon dislocations~\cite{Yazyev:2010p3429} revealed the possibility of forming bands around zero energy when the dislocations are close to each other, accounting for the \ldos peaks at finite energies.

Motivated by the STS measurement results, we aim at extending the theoretical knowledge of extended defect structures by analyzing the electronic structure of amorphous tilt grain boundaries in graphene, expecting our results to be directly applicable to measurements on the surface of graphite. Our approach is based on considering the relaxed boundary of misaligned grains of graphene, and the results should be of direct relevance to the structures found along the grain boundaries as seen on the surface of graphite.~\cite{Zhou:2006p3439} We find that the disordered structures formed at the relaxed boundary between two differently oriented grains can have enhanced \ldos at zero, or at finite energies. These features result from narrow bands (localized states) that can form both near and away from zero energy. We also discuss  grain boundary models derived from dislocation arrays, considering  dislocation cores that are different from the simplest pentagon-heptagon structure.~\cite{Carpio:2008p3430} These can lead to \ldos enhancement at zero energy as seen in the STM measurements, that are not seen in the pentagon-hexagon model of Ref.~\onlinecite{Yazyev:2010p3429}. Finally, we do identify a  special limit where the zero modes of the  low energy continuum theory of dislocated graphene precisely agrees with  tight-binding model results.  Intriguingly, this theory predicts the appearance of localized zero energy states in an array of well separated dislocations, in contrast to the results of the first principles calculations of Ref.~\onlinecite{Yazyev:2010p3429}.

This paper is organized as follows. In Section~\ref{sec:dislocation-as-base} we review the \ldos of dislocations, considering two defects at different distances, as well as the isolated case. Next in Section~\ref{sec:cont-model-disl} we use the continuum theory of graphene to explain the density of states and predict the zero energy peak in an array of dislocations. In Section~\ref{sec:electr-tight-bind} we present our study of the tight-binding model of relaxed tilt grain boundaries in graphene, with a variety of opening angles. We close with discussion and conclusions.

\section{Dislocations in graphene as base of grain boundary models}
\label{sec:dislocation-as-base}

Dislocation models of grain boundaries rely on the fact that an array of dislocations with same Burgers vectors produces a boundary line between two crystal domains of different lattice orientations.~\cite{Burgers:1940p3435,Bragg:1940p3436,Read:1950p3434} In this Section we make observations relevant to such models in graphene, inspired by the recent STM experiments.~\cite{Cervenka:2008p2547}

\subsection{Graphene dislocations in tight-binding}
\label{sec:graph-disl}
%%%%%%%%%%%%%%%%%%%%%%%%%%%%%%%%%%%%%%%
\begin{figure*}
%  \begin{widetext}  
  \centering
\includegraphics[width=1\textwidth]{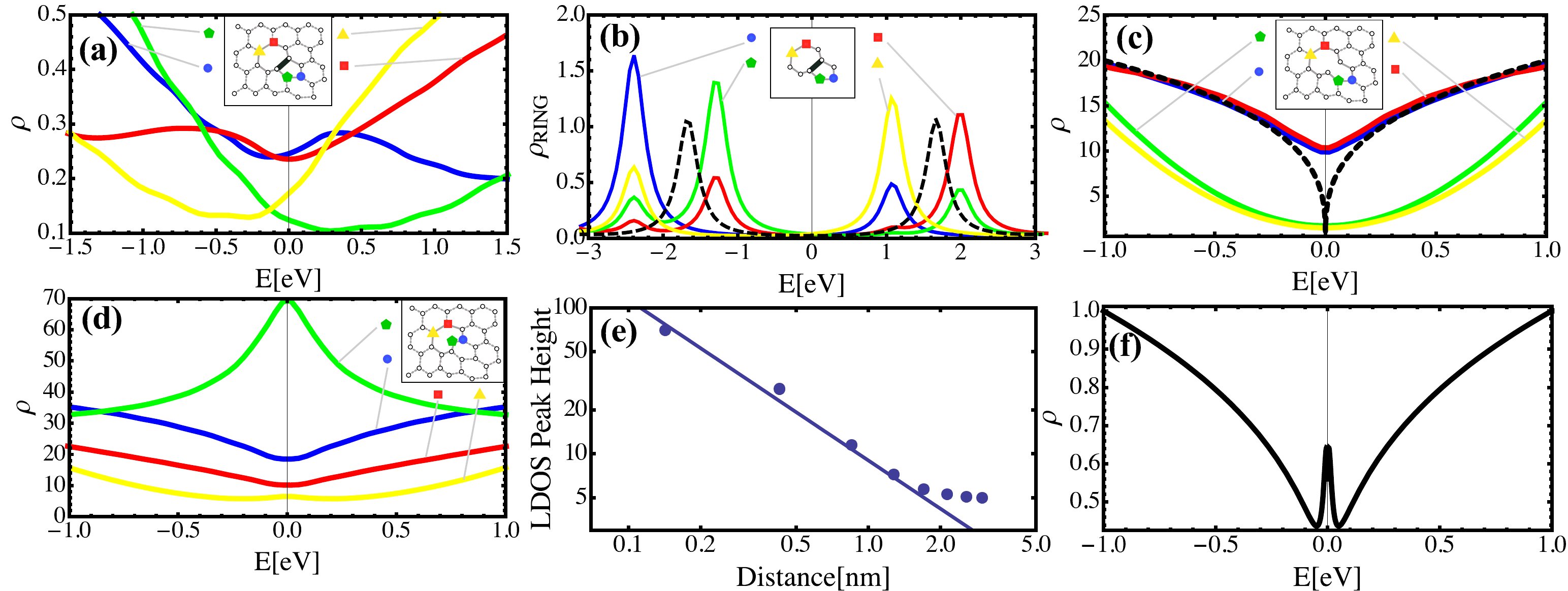}
\caption{The \ldos of graphene dislocations from tight-binding and continuum theory. (a) \ldos of representative atoms of the ``PH'' type dislocation core (inset, see Section~\ref{sec:graph-disl}). As discussed in section IIB, the weight is shifted due to the A-A bond (thick in the inset), compared to the symmetric curves in (c)
obtained by switching off the A-A bond that are consistent with continuum theory . (b) The influence of the A-A bond on an isolated ring: the lattice case (a) can be viewed as a broadened version. (c) ``PH'' core \ldos without the AA bond. Dislocation topology effects from continuum theory (Section~\ref{sec:cont-model-disl}, dashed black curve) are prominent. (Finite \ldos at $E=0$ is a finite size effect.) (d) \ldos of representative atoms in ``OCT'' core type. (e) The height of \ldos peak in (d) (pentagon, green) falls of like a power-law with distance from defect core. This holds also for \ldos features in (a). (f) The continuum defect topology prediction of \ldos (in patch of radius $\delta=0.1$, Section~\ref{sec:cont-model-disl}) in a dislocation array, with zero modes.}
\label{fig:1}
 % \end{widetext}
\end{figure*}
%%%%%%%%%%%%%%%%%%%%%%%%%%%%%%%%%%%%%%%

Simple dislocation cores in graphene come in two shapes that we label ``PH core'' and ``OCT core'' (cf. Fig.~\ref{fig:1}(a),(d)), both of which were shown to be stable lattice configurations.~\cite{Hashimoto:2004p575,Carpio:2008p3430} Geometrically, the two possibilities arise because the Bravais lattice has two atoms (separated by $\bd$) in the unit-cell, so that there are two inequivalent mutual configurations of $\bd$ and the Burgers vector $\bb$. The \ldos at the atoms forming the core has been considered,~\cite{Tamura:1997p1095,Carpio:2008p3430} revealing a sharp peak at zero energy in case of the ``OCT'' core, due to the undercoordinated atom. Note that even if only $\pi-$orbitals are considered, the single excess atom in one sublattice carries an \ldos peak and the accompanying (locally unbalanced) magnetic moment in the presence of interactions.~\cite{Carpio:2008p3430,LopezSancho:2008p219,Pereira:2006p220,Vozmediano:2006p3432} Alternatively, the ``OCT'' core can be viewed as a piece of a zigzag graphene edge of minimal length of one atom embedded in the graphene bulk, leading to same conclusions about its \ldos features.~\cite{Vozmediano:2005p2562,Lee:2005p1143,Nakada:1996p717,Pisani:2007p682,Son:2006p605,Wunsch:2008p357,Palacios:2008p3440}

Ref.~\onlinecite{Yazyev:2010p3429} considers only ``PH'' type cores as building blocks of grain boundaries, and such models have been proposed in earlier graphite STM measurements.~\cite{Simonis:2002p1044,Yoon:2001p549} However, the zigzag oriented grain boundary model of Ref.~\onlinecite{Cervenka:2009p2555}, as well as simple geometrical considerations we present above, both show that the arrays of ``OCT'' type dislocations should not be disregarded in real materials, even if they are more energetically costly than the ``PH'' type.

The inclusion of the ``OCT'' dislocations can be important for explaining the observed \ldos peaks at zero energy in the measurements of Ref.~\onlinecite{Cervenka:2009p2555}. The set of grain boundaries considered there shows that such \ldos features are found only when the defect cores are well separated (i.e. the grain boundary angle is small). One might assume that when the defects are closer to each other, the zero energy states hybridize and move to finite energies. We have however found that the localized zero energy modes are robust even when the defects are brought next to each other, which would be the case in a grain boundary with maximal opening angle.

Our analysis was done by considering the \ldos of defects set inside a $75\mathrm{x}75$ unit-cell sized graphene patch tight-binding model, with twisted periodic boundary conditions in both directions. The special boundary conditions enable the system under consideration to actually be a periodic, $10\mathrm{x}10$ sized arrangement with the graphene patches as unit-cells, thereby leading to a tenfold increase in linear system size and a correspondingly denser energy spectrum ${E_n}$ (with corresponding eigenfunctions $\psi_n$), from which the \ldos $\rho(i,E)$ at site $i$ and energy $E$ follows in a standard way:
\begin{equation}
  \label{eq:1}
  \rho(i,E)=\frac{1}{\pi}\sum\limits_n|\psi_n(i)|^2\mathrm{Im}{\frac{1}{E-E_n+i\varepsilon}}.
\end{equation}
We applied a small broadening $\varepsilon\approx 20$ meV of levels into a Lorentzian shape, which is both expected to exist in the material and leads to smoothing of the finite size effects in the \ldos. We introduce the defects by inserting a line of extra atoms, thereby creating a defect---anti-defect pair at a maximum separation of half the graphene patch size. By adding an additional line of atoms, we can study the \ldos of two defect cores close to each other, isolated from their anti-defects. The Hamiltonian is of the single particle spin degenerate tight-binding graphene:
\begin{equation}
  \label{eq:2}
  H=-\sum\limits_{<ij>}t_{ij}(c^\dagger_ic_j+H.c.),
\end{equation}
with the hopping constant $t=2.7$ eV. When choosing the nearest neighbor pairs in Eq.~\eqref{eq:2}, we retain the topology of the honeycomb lattice, which is violated only at a single atom in the ``OCT'' dislocation case. The \ldos turns out to be robust to relaxation of bond lengths, so that the results for $t_{ij}=t$ are representative.

Our calculation shows that the \ldos at the dislocation cores is insensitive to the distance between the dislocations, in particular the \ldos peak at zero energy in the ``OCT'' type core system stays pinned and does not hybridize when the defects are brought close to each other to minimal distance of few lattice constants.

The results of the tight-binding model presented in this section show that the characteristic features of the dislocation \ldos, notably the zero energy peak of the ``OCT'' core fall-off with distance from the core as a power law (Fig.~\ref{fig:1}(e)). This is the expected behavior according to low energy continuum models of graphene (see Section~\ref{sec:cont-model-disl}), and also argued for in the case of cracks in graphene in Ref.~\onlinecite{Vozmediano:2005p2562}.

The STS measurements of Ref.~\onlinecite{Cervenka:2009p2555}, achieving atomic resolution, however show an exponential fall-off of \ldos features with the distance from the prominent defect centers, even for defects far from each other. This discrepancy might be due to subtle shortcomings of substituting a simplified single graphene sheet for the top layer of graphite; however, another explanation could be the presence of stronger disorder. The fact that the single atom resolution along the grain boundary is lost in patches of several lattice constants across also indicates that the grain boundary might contain more disorder than an array of simple dislocations. This presents additional motivation for our study of amorphous tilt grain boundaries presented in Section~\ref{sec:electr-tight-bind}, in place of the coherent ones studied in Refs.~\onlinecite{Yazyev:2010p3429,Yoon:2001p549,Simonis:2002p1044}.

\subsection{Continuum model of dislocations}
\label{sec:cont-model-disl}

It is interesting and fundamental to approach the description of grain boundaries by considering an analytical model. In this Section, we describe the results of such a continuum model, finding conditional agreement with the tight-binding results. We then proceed to use the theory for describing the \ldos of an array of dislocations, and find a surprising prediction of localized modes at zero energy. Even if the continuum theory prediction fails in a more realistic model (as Ref.~\onlinecite{Yazyev:2010p3429} suggests), we find it a fundamental step in understanding the system.

The continuum description of the topological effect of dislocations is based on the description of the defect as a translation by the Burgers vector $\bb$ of the wavefunction of the ideal crystal, upon encircling the defect core. The model is therefore akin to an Aharonov-Bohm (AB) effect, except that it does not break time reversal symmetry. The details of this model are derived in Ref. ~\onlinecite{Mesaros:2009p1905}, and here we start from the Hamiltonian in the form of the standard graphene Dirac equation, coupled to a dislocation gauge field,
\begin{equation}
  \label{eq:4}
  H_{disl}=-i\hbar v_F \tau_0\otimes\vsig\cdot(\grad-i\A),
\end{equation}
where the dislocation gauge field $\A$ (in fixed gauge) produces the correct pseudoflux of the translation holonomy $\oint \A\cdot d{\bm x}\equiv (\bK\cdot\bb)\tau_3=2\pi d\;\tau_3$, e.g. $A_\varphi=\frac{(\bK\cdot\bb)}{2\pi r}\tau_3=\frac{d}{r}\tau_3$, where $r$ and $\varphi$ are the standard polar coordinates. The Burgers vector is encoded in the dislocation pseudoflux $d$ which has only three inequivalent values $\{0,\frac{1}{3},-\frac{1}{3}\}\equiv \{0,-\frac{1}{3},-\frac{2}{3}\}$, opposite at the two Fermi points.~\cite{Mesaros:2009p1905} We label a Fermi wavevector by $\bK$ (and the other Fermi point is at $-\bK$), $\tau$ matrices mix the two Dirac points, the $\sigma$ matrices act on the $A/B$ sublattice, and we use the four component spinor $\Psi(\br)\equiv (\Psi_{\bKp A},\ \Psi_{\bKp B},\ \Psi_{\bKm B},\ -\Psi_{\bKm A})^T$.

An important property of the translation operator, and consequently the $\A$ gauge field, is that it does not mix the Fermi points, so that we can consider them separately. Therefore our model is based on a single valley Dirac equation in the AB field of flux $d\in\{-\frac{1}{3},-\frac{2}{3}\}$,
\begin{equation}
  \label{eq:5}
    H^d_+=-i\hbar v_F \vsig\cdot(\grad-i \frac{d}{r}\vef).
\end{equation}
The other valley experiences the complementary flux $-1-d$, i.e. $H^d_-=H^{-1-d}_+$. We have chosen the values of $d$ such to conform to the practice of AB flux being the fractional flux part.

To test this theory, we find that the \ldos in a patch of radius $\delta$ covering the defect behaves as,
\begin{equation}
  \label{eq:3}
  \rho(\delta,E)\sim \delta^{4/3}|E|^{1/3}.
\end{equation}
This actually agrees with the tight-binding model results in the limit where the bipartiteness of the honeycomb lattice is not broken by the defect, see Fig.~\ref{fig:1}(c). This is precisely  the limit  where we expect  that the effects of the global topology in the hopping network become dominant. This condition  can be realized  in
principle for both cores pending their  'chemistry'. In the ``OCT'' case the undercoordinated atom appears as an intruder, but otherwise the bipartitness and topology of the ideal lattice are preserved. In the ``PH'' core case, the A-A bond (inset of Fig.~\ref{fig:1}a) spoils the hopping bipartiteness  when it supports a finite hopping.  It interferes with the purely topological effect of the dislocation, and introduces asymmetric features in the \ldos   (Fig.~\ref{fig:1}(a)) while the powerlaw behavior expected from the continuum limit is recovered when the bond is switched off (Fig.~\ref{fig:1}(c)). The origin of the asymmetric features is clearly identified by considering the \ldos of an isolated $10$ atom ring which is turned into a pentagon---heptagon structure by switching on the A-A bond  (Fig.~\ref{fig:1}(b)): the 
lattice results (Fig.~\ref{fig:1}(a)) can be viewed as the 'molecular' states of Fig.~\ref{fig:1}(b) turning into broadened, resonant impurity bound states. 

We now outline the calculation leading to Eq.~\eqref{eq:3}, which is also fundamental for understanding the prediction for a dislocation array. The eigenfunctions of Eq.~\eqref{eq:5} are found by separating the angle, and we find for energy $E=\epsilon\hbar v_F\lambda$ ($\epsilon=\pm 1$ and $\lambda>0$):
\begin{equation}
  \label{eq:6}
  \Psi^s_E(\br)=\sum_{m\in\mathbb{Z}}e^{i m\varphi}
  \begin{pmatrix}
    e^{-i\varphi} u^s_m(r) \\
    \epsilon i v^s_m(r)
  \end{pmatrix},
\end{equation}
where the sign $s=+,-$ labels two linearly independent solutions $\psi^s_m\equiv (u^s_m(r), v^s_m(r))^T=(J_{s(m-1-d)}(\lambda r), J_{s(m-d)}(\lambda r))^T$, and $J_q$ is the Bessel function of order $q$ (note that $q\notin\mathbb{Z}$). The total angular momentum in channel $m$ is $j=m-1/2$, and we see that the presence of dislocation shifts it $j\rightarrow j-d$. Normalizability allows exclusively $\psi^+_m$ for $m>0$, and $\psi^-_m$ for $m<0$, and both for $m=0$. The Hamiltonian Eq.~\eqref{eq:5} is actually not self-adjoint, so that there is additional physical input needed regarding the wavefunction boundary condition at the singular point at the defect. At this point the theory becomes sensitive to the 'UV', that is the microscopic details at the lattice cut-off. In the field theoretical derivation that follows this
UV regularization is kept as featureless as possible. However, we find from the explicit tight binding description that the 'chemistry' of the core structure does matter. Even without the A-A bond, which spoils the global topology, the PH core cannot be represented by the continuum theory (Fig.~\ref{fig:1}(c)).  
However, a key result of this paper is that the OCT dislocation core (Fig.~\ref{fig:1}(d),(e)) is compatible with the continuum theory zero mode (Fig.~\ref{fig:1}(f)).

The application of the standard theory of self-adjoint extensions (SAE)\cite{Weidmann,Thaller,Fulop:2007p3442,Jackiw:1991p1161} prescribes  that the coefficients of the linear combination $N_+\psi^+_0+N_-\psi^-_0$ in channel $m=0$ determine the additional physical parameter $\chi\in[0,2\pi)$ through
\begin{equation}
  \label{eq:7}
  N_+/N_-=\cot{(\chi/2)}.
\end{equation}
The channel $m=0$ actually contains normalized spinors $\psi^{+/-}_m$ which have diverging components on the sublattice $A/B$ respectively, and the ratio of these divergences is set by the particular SAE through the value of $\chi$.

We can now evaluate the \ldos in a patch of radius $\delta$,~\cite{Lammert:2000p541}
\begin{align}
  \label{eq:8}
  \rho(\delta,E)&=2\pi\int_0^\delta r dr\sum_{\epsilon,\lambda}\sum_m|\psi_m(\lambda r)|^2 \delta(E-E_\lambda)\sim\\
  &\sim \int \lambda d\lambda\int_0^\delta r dr \left(r\lambda\right)^{2q} \delta(E-\hbar v_F\lambda)\sim\\
  &\sim \delta^{2q+2}|E|^{2q+1},
\end{align}
where in the second line we have used the Bessel function density of states $\sum_\lambda\rightarrow\int\lambda d\lambda$, and evaluated the small argument (i.e. $\lambda r\ll 1$) expansion of the Bessel functions of order $q$. The leading contribution comes from the diverging components of $\psi_0$, where $q=-1/3,-2/3$ (this holds for both Fermi points, and both dislocation classes).

The value of $q=-2/3$ generates an unphysical divergence $\rho\sim 1/|E|^{1/3}$. We therefore have to choose the SAE which removes the offending part of $\psi_0$, and this turns out to be $\chi=\pi$ and $\chi=0$, for $d=-1/3$ and $d=-2/3$, respectively. Note that at the second Fermi point, we have to switch the values, so that $\chi=0,\pi$ for $d=-1/3,-2/3$. The surviving components in $\psi_0$ with $q=-1/3$ yield the advertised patch \ldos of Eq.~\eqref{eq:3}.

\subsection{Continuum model of dislocation arrays}
\label{sec:cont-model-disl-1}
%%%%%%%%%%%%%%%%%%%%%%%%%%%%%%%%%%%%%%%
\begin{figure*}
  \centering
\includegraphics[width=1\textwidth]{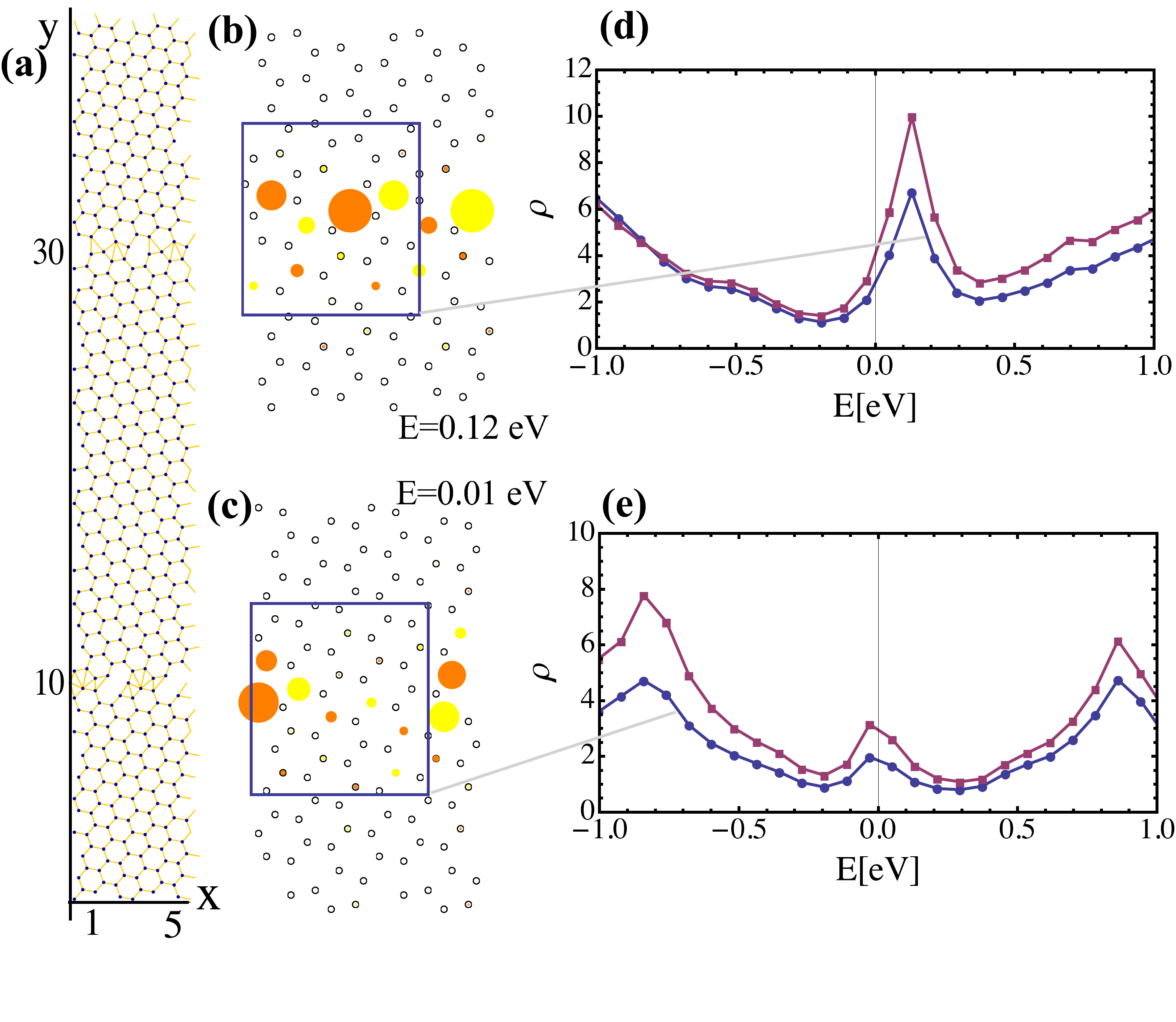}
\caption{The \ldos and states of tight-binding amorphous tilt grain boundary: example of armchair type with medium opening angle. (a) The system with hoppings (of different strength in calculation) included. (b),(c) Zoom-in of the boundaries, including the wavefunctions at $k_x=0$ at energy of the \ldos peaks: size of colored dots is the amplitude, orange (dark grey) and yellow (light grey) denote opposite sign. (d),(e) The \ldos of two grain boundaries, averaged within square patches as marked in (a).}
\label{fig:2}
\end{figure*}
%%%%%%%%%%%%%%%%%%%%%%%%%%%%%%%%%%%%%%%

Once we know the details of the continuum description of a graphene dislocation derived in the previous subsection, we can ask the question: what happens in an array of such defects? As we have shown, the SAE of the continuum Hamiltonian of Eq.~\eqref{eq:5} is fixed by the allowed values of $\chi=0,\pi$. It turns out that precisely these special values of $\chi$ allow the Hamiltonian to have localized states at zero energy. This leads to a peak at zero energy which is absent from the gapless, cusp shaped \ldos of the finite energy wavefunctions in Eq.~\eqref{eq:3}.

It is well known, that Hamiltonians with singular potentials (e.g. AB flux,~\cite{Persson:2006p260} Coulomb potential,~\cite{Fulop:2007p3442} delta function potential~\cite{Jackiw:1991p1161}), once they are made Hermitian through a SAE, can exhibit finite or zero energy bound states, even if the original Hamiltonian was scale-free.

Our system is represented by two copies (two Fermi points) of a two-component, two dimensional spinor in the presence of a pseudomagnetic solenoid with flux $d\in\{-1/3,-2/3\}$. The problem of a spinful two dimensional particle moving in an arbitrary magnetic field, both non-relativistic (Pauli) and relativistic (Dirac), has originally been considered by Aharonov and Casher,~\cite{Aharonov:1979p277} who found that the number of flux quanta give the number of zero-energy states of the particle. In Ref.~\onlinecite{Persson:2006p260}, the Dirac particle in the presence of multiple AB solenoids is considered, so we can here directly use those results concerning the zero modes of the Dirac Hamiltonian of the form Eq.~\eqref{eq:5}.

Let us describe the relevant calculation, following Refs.~\onlinecite{Persson:2006p260,Aharonov:1979p277} closely. The fact that $\chi=0,\pi$ is the key ingredient: as we have seen in Section~\ref{sec:cont-model-disl}, at these values the divergence of the wavefunction is allowed in only one of the spinor components. This means that the SAE imposed boundary condition on the wavefunction does not mix the two components, i.e. sublattices. For zero energy, the eigenproblem of $H^+_d$ also decouples the sublattices. Going to complex coordinates $z=x+i y$, $A=A_x+i A_y$, and using the scalar potential $\Phi(z)=-\sum^n_i d_i \log{|z-z_i|}$ of the AB gauge potential of dislocations $d_i$ at positions $z_i$ (i.e. $\partial_z\Phi(z)=A$), we can solve for the two sublattices separately: $\partial_{z^*}(e^{-\Phi} u)=0$ and $\partial_z(e^\Phi v)=0$. The Dirac equation now tells us that $e^{-\Phi} u$ ($e^\Phi v$) is an analytic (antianalytic) function outside the singular points $z_i$. For $\chi=\pi$, $u$ cannot have singularities according to Eq.~\eqref{eq:7}. Taking into account the behavior $e^{-\Phi}\sim |z|^{\phi},\;|z|\rightarrow\infty$, with $\phi=-\sum^n_i d_i$ the total pseudoflux, it follows that $u$ can be a polynomial of $z$ of order at most $\{-\phi\}-1$, where $\{\}$ is the lower integer part. There are $\{-\phi\}$ linearly independent such polynomials. In the case $\chi=0$, $v$ is not singular so that $e^\Phi v$ vanishes at the defects. $v$ is a polynomial in $z^*$ of degree $\{\phi\}-1$, with $n$ zeros, and there are $\{\phi-n\}$ of them.

Collecting the results, there are $\{|\phi-n|\}$ ($\{|\phi|\}$) zero modes in the case $\chi=0$ ($\chi=\pi$) for the single Fermi point system.  Note that we assumed the same value of $\chi$ for each dislocation $d_i$, so that the result holds only in the case of all dislocations having equivalent Burgers vectors, which is the case of a grain boundary! The two Fermi points contribute independently to the number of zero modes, and since $\chi$ and $d_i$ are reversed between them, we get a total of
\begin{equation}
  \label{eq:9}
D=\{|2\phi-n|\}+\{|\phi|\}, \text{ with } \phi=\frac{n}{3},
\end{equation}
zero modes in graphene with an array of $n$ dislocations having the same Burgers vector (of whichever non-trivial class $d$). This number takes the values $D=2,2,2,4,4,4,6\ldots$, starting at $n=4$ and onwards. More precisely, we get $D=2 \{n/3\}$, so that $D$ scales with the system size, i.e. $D\sim\frac{2}{3}n$ in the thermodynamic limit.

The zero energy modes are localized at the defects and have a power-law shape. To answer the question of whether they are observable, we look at how the \ldos in a patch of radius $\delta$ scales in comparison to the \ldos contribution of the finite energy states Eq.~\eqref{eq:3}. Near the defect at $z_i$, at one Fermi point the $u$ spinor component (sublattice A) scales as $|z-z_i|^{-d}(z-z_i)^p$, where $p\leq \{-\phi\}-1$ and the value of $d$ is $-1/3$. This gives a contribution $\rho(\delta)\sim \delta^{2p-2d+2}$. The same sublattice at the other Fermi point contributes through $v\sim |z-z_i|^{1+d}(z-z_i)^t$, with $t\leq \{n-\phi\}-1$, giving $\rho(\delta)\sim \delta^{2t-2d}$. In the case of the opposite defect type, we get the same scaling, but on sublattice B. The leading contribution in the \ldos comes from the minimal values of $p=0$ and $t=0$, giving one mode at the defect with the \ldos
\begin{equation}
  \label{eq:11}
  \rho^{(0)}(\delta)\sim \delta^{2/3}.
\end{equation}
The scaling shows that the zero mode contribution $\rho^{(0)}$ is more favourable than the finite energy contribution $\rho(\delta,E)\sim \delta^{4/3}E^{1/3}$ at smaller $\delta$, because the strongest zero modes are more localized than all the finite energy wavefunctions.

Since in this Section we dealt with a Dirac particle, it is interesting to consider the number of zero modes through the Atiyah---Singer theorem: In graphene with disclinations this was already analyzed in Ref.~\onlinecite{Pachos:2006p100} by using the defect gauge field of a disclination. There it was shown that the number of zero modes is proportional to the Euler characteristic of the manifold. The key to the application of the theorem is that graphene with disclinations can form compact manifolds, e.g. the fullerene molecule, so that the mapping of the lattice and hopping topology onto a compact continuum manifold is correct, leaving the low energy Dirac particle description valid. For the case of dislocations however there is no such possibility. The exception is the map from the dislocated lattice onto the torus (i.e. the plane with periodic boundary conditions), but this is possible only when for every dislocation there is an anti-dislocation. In that case, the theorem predicts no zero modes, in accordance with our present calculation.

\section{Electronic tight-binding model of relaxed symmetric amorphous grain boundaries in graphene}
\label{sec:electr-tight-bind}
%%%%%%%%%%%%%%%%%%%%%%%%%%%%%%%%%%%%%%%
\begin{figure*}
  \centering
\includegraphics[width=1\textwidth]{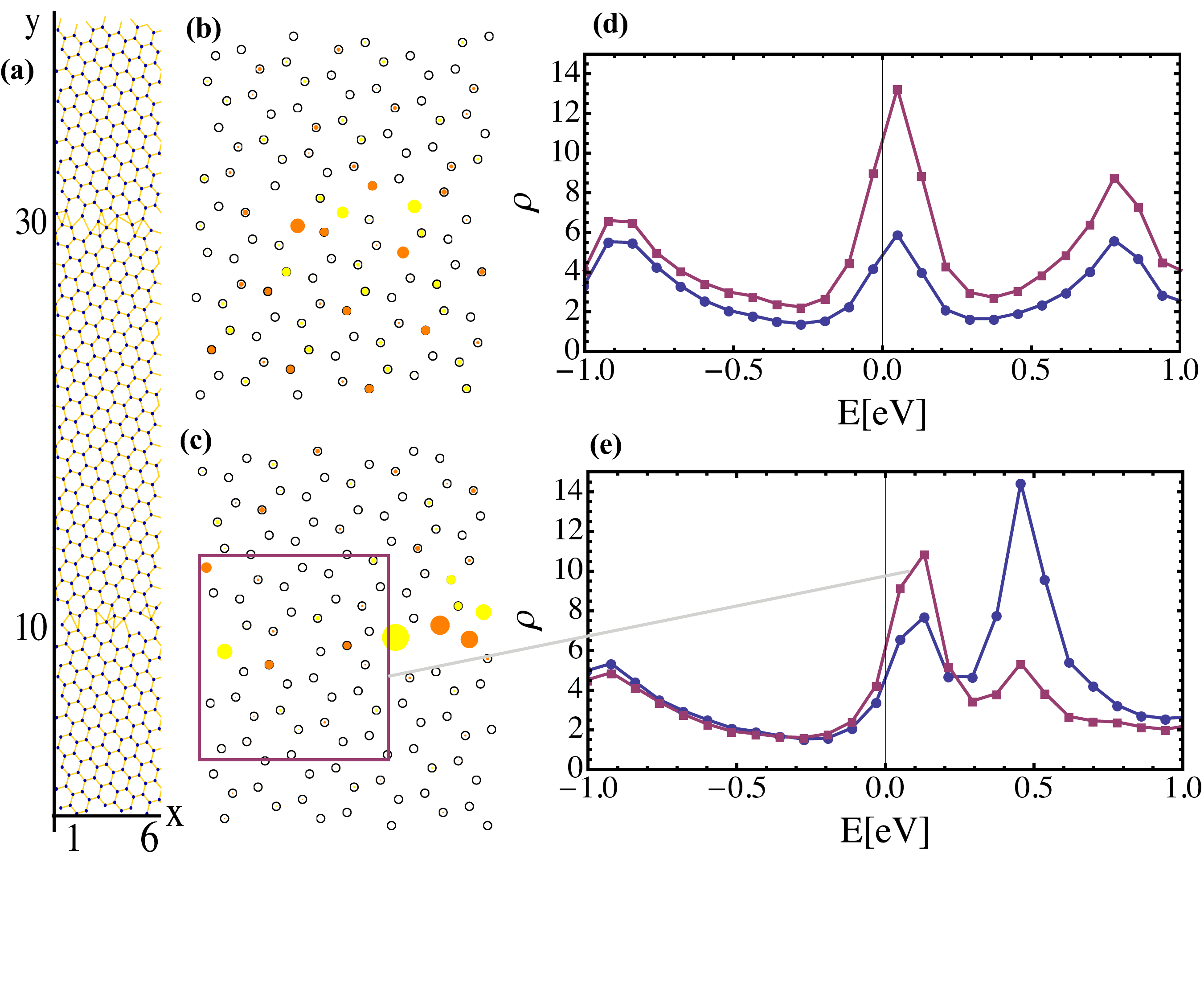}
\caption{The \ldos and states of tight-binding amorphous tilt grain boundary: example of zigzag type of medium opening angle. (a) The system with hoppings (of different strength in calculation) included. (b),(c) Zoom-in of the boundaries, including the wavefunction at $k_x=0$ and $E=012$ eV: size of colored dots is the amplitude, orange (dark grey) and yellow (light grey) denote opposite sign. (d),(e) The \ldos of two grain boundaries, averaged within square patches as marked in (a).}
\label{fig:3}
\end{figure*}
%%%%%%%%%%%%%%%%%%%%%%%%%%%%%%%%%%%%%%%
\subsection{The Method}
\label{sec:method}

We consider two misoriented graphene grains of same width, confined in a periodic box of width $L_y$ and length $L_x$. The nominal box boundaries at $y=0,L_y$ are positioned through the middle of the width of the ``first'' grain. The second grain is generated in the middle half of the box (from $L_y/4$ and $3L_y/4$) and both grains are periodic with the box in the $x$ direction. The boundaries at $x=0$ and $x=L_x$ have twisted PBC so that the system is a periodic crystal of length $N*L_x$ (we set $N=18$), with momenta $k_x=2\pi/(N L_x)$. The lattice of each grain is generated from its center, and terminated at the grain boundaries. The boundaries are symmetric, but in general the two grain boundaries do not have the same structure because the two grains have different centers of inversion symmetry.

The allowed values for the grain's orientation $\theta_i$ follow from the constraint of its periodicity with the box along the $x$ direction. If $\vLp=a_i\veone+b_i\vetwo$ is the vector in the basis of the graphene Bravais lattice which is to be wrapped along the $x$ box direction, the constraint is
\begin{equation*}
  \cos{(\theta_i)}=\frac{a_i+b_i/2}{a_i^2+b_i^2+a_ib_i/2},
\end{equation*}
as also explicated in Ref.~\onlinecite{Coffman:2008p3450}. If we allow slight strain in the grain, the number of available orientations can be enlarged.~\cite{Coffman:2008p3450} The grain opening angle $\theta=\theta_1-\theta_2=2\theta_1$ spans the entire $[0^\circ,30^\circ]$ range (for both zigzag and armchair type~\cite{Cervenka:2009p2555,Yazyev:2010p3429}).

We relax the atoms in the system at zero temperature using the molecular dynamics method, where the interatomic potential for carbon is taken in the Tersoff---Brenner form.~\cite{Tersoff:1988p3448,Brenner:1990p3449} The potential between atoms $i,j$ at distance $r_{ij}$ is
\begin{align*}
  V(r_{ij})&=V_R(r_{ij})-\bar{B}_{ij}V_A(r_{ij})\\
  V_R(r)&=\frac{D}{S-1}e^{-\sqrt{2S}\beta(r-R)}f(r)\\
    V_A(r)&=\frac{D S}{S-1}e^{-\sqrt{2/S}\beta(r-R)}f(r),
  \end{align*}
  with $f(r)$ the smoothing function
\begin{equation*}
f(r)=  \begin{cases}
  1,&r<R_1\\
  \frac{1}{2}\left(1+\cos{\left[\frac{(r-R_1)\pi}{R_2-R_1}\right]}\right),&R_1<r<R_2\\
  0,&r>R_2.
  \end{cases}
\end{equation*}
The effect of bond angles is encoded in $\bar{B}_{ij}=1/2(B_{ij}+B_{ji})$, with
\begin{equation*}
  B_{ij}=\left(1+\sum\limits_{k\neq i,j}G(\theta_{ijk})f(r_{ik})\right)^{-\delta},
\end{equation*}
where $\theta_{ijk}$ is the angle between the $i-j$ and $i-k$ bonds, and the function $G$ is
\begin{equation*}
  G(\theta)=a_0\left(1+\frac{c_0^2}{d_0^2}-\frac{c_0^2}{d_)^2+(1+\cos{(\theta)})^2}\right).
\end{equation*}
The ground state of this non-spherically symmetric potential is the graphene honeycomb lattice, when the parameters are chosen as in Ref.~\onlinecite{Brenner:1990p3449}: $D=6$ eV, $R=0.139$ nm, $\beta=21\;nm^{-1}$, $S=1.22$, $\delta=0.5$, $a_0=0.00020813$, $c_0=330$, and $d_0=3.5$. The smoothing cutoffs are chosen to include the nearest neighbor atoms, $R_1=0.17\;nm$ and $R_2=0.2\;nm$.

When the lattice is formed, we consider a tight-binding model for electrons, of the form Eq.~\eqref{eq:2}. The hopping constants $t_{ij}$ are taken to fall-off exponentially, and fitted so that $t_{ij}$ for the nearest neighbor distance $|\Delta|$ is $t=2.7$ eV, and for the next-nearest neighbor distance $\sqrt{3}|\Delta|$ it is $t'=0.1$ eV, in accordance with accepted values for graphene.~\cite{PhysRevB.66.035412} Finally, we extract the energy bands $E(k_x)$, the wavefunctions $\psi_E(i)$ and the \ldos $\rho(i,E)$.

\subsection{Summary of Results}
\label{sec:analysis-results}
%%%%%%%%%%%%%%%%%%%%%%%%%%%%%%%%%%%%%%%
\begin{figure*}
  \centering
\includegraphics[width=1\textwidth]{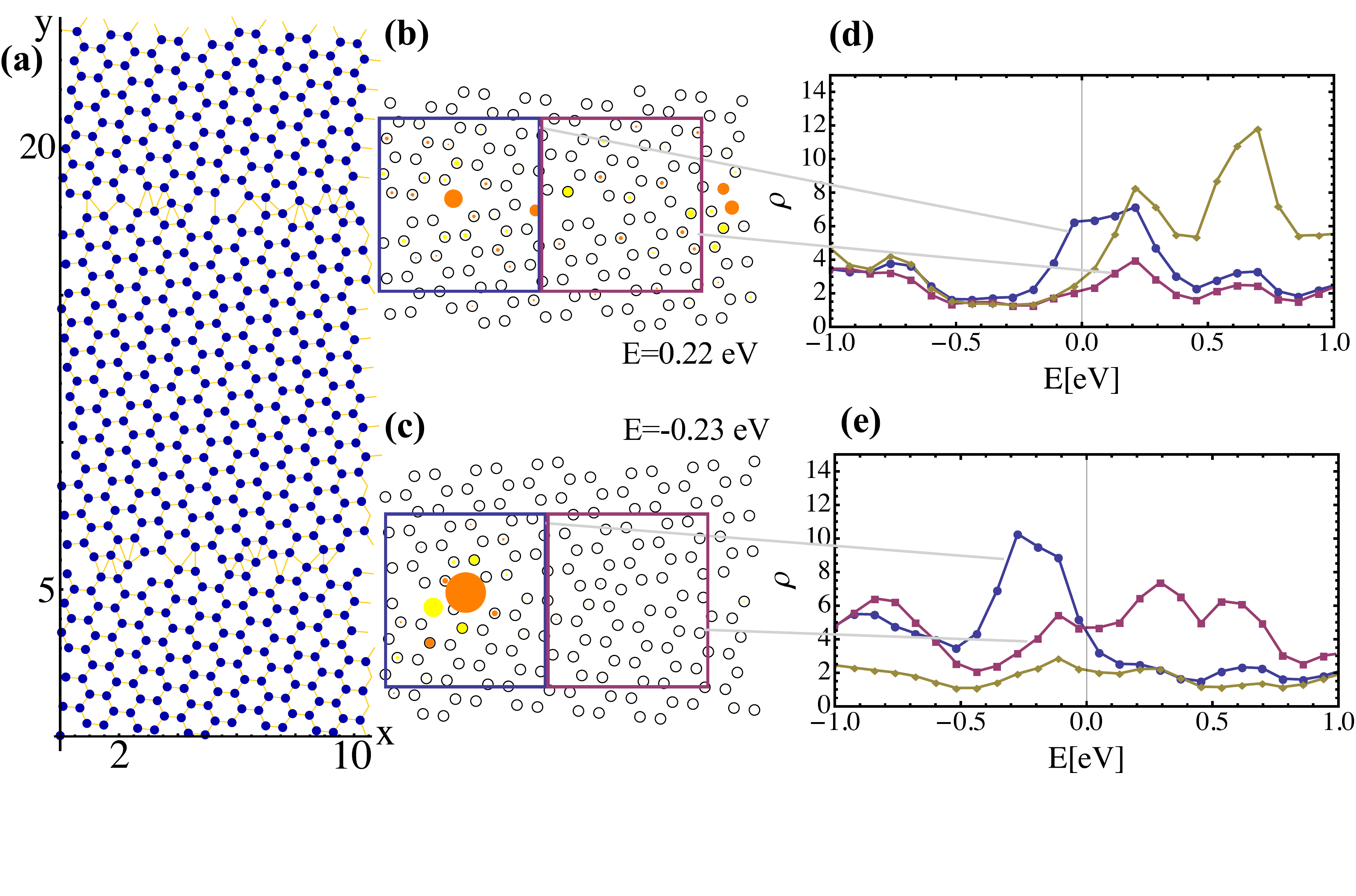}
\caption{The \ldos and states of tight-binding amorphous tilt grain boundary: example of armchair type of small opening angle. (a) The system with hoppings (of different strength in calculation) included. (b),(c) Zoom-in of the boundaries, including the wavefunctions at $k_x=0$ at $E$ of the peaks: size of colored dots is the amplitude, orange (dark grey) and yellow (light grey) denote opposite sign. (d),(e) The \ldos of two grain boundaries, averaged within square patches as marked in (a).}
\label{fig:4}
\end{figure*}
%%%%%%%%%%%%%%%%%%%%%%%%%%%%%%%%%%%%%%%

We analyze in detail a number of grain boundaries of both zigzag and armchair type,~\cite{Cervenka:2009p2555,Yazyev:2010p3429} covering the entire range of opening angles by varying the box size: $2.6a<L_x<16.1a$ with $a=0.246$ nm the graphene Bravais lattice constant.

In summary, we find that the \ldos along the grain boundaries, averaged in square patches of size $4a$ and considered in the low energy regime of $|E|<1$ eV, shows three typical behaviors:
\begin{description}
\item[(i)] A peak at very small energy, $|E|<0.05$ eV
\item[(ii)] Two peaks at nearly opposite energies, at around $0.3\;\text{eV}<|E|<0.5$ eV
\item[(iii)] Just one peak, at an energy $0.3\;\text{eV}<|E|<0.5$ eV.
\end{description}

Focusing on case (i), we have determined that the lowest energy wavefunctions are sometimes localized on structures that resemble short zigzag edge segments (i.e. of length $2a$). This however occurs also in armchair type boundaries, but of course then the short zigzag segment is tilted away from the grain boundary line, the $x$ axis. In some systems however, the zero energy peak is associated with overcoordinated atoms, having even five neighbours.

We find that clear examples of case (ii) mostly appear at high opening angles (i.e. small $L_x$), where the strong \ldos signal spans the entire grain boundary. There are also just a few energy bands with $|E|<1$ eV, so it is easy to identify that the \ldos peaks are due to van Hove singularities, in accordance with the findings of Ref.~\onlinecite{Yazyev:2010p3429} for large opening angles.

The case (iii) we find is strongly correlated with carbon atoms that were annealed into a position with four neighbors, meaning that four atoms are within the $|\bd|$ distance, distributed roughly evenly around the central atom.

Since the \ldos behavior of case (ii) has already been identified in Ref.~\onlinecite{Yazyev:2010p3429}, we illustrate the occurrence of cases (i) and (iii) through typical examples Figs.~\ref{fig:2},\ref{fig:3},\ref{fig:4}.

Finally, we note that typically there is one localized region within the box that has atoms with high \ldos values, i.e. one can say that there is one prominent ``defect'' within one $L_x$ long unit-cell of the entire grain boundary. This means that the periodicity of grain boundary calculated from the opening angle corresponds to the periodicity of prominent defect structures, even in our case of amorphous boundaries.~\cite{Cervenka:2009p2555} There are rare special cases where our box has an accidental symmetry so that the defect structures along the boundary repeat twice within the box length $L_x$, effectively halving $L_x$ and $\theta$.

\section{Discussion and Conclusions}

We have analyzed the electronic structure of a variety of grain boundaries that can form in graphene. Quite likely the grain boundaries that are formed 
spontaneously in graphite, and that are best characterized experimentally, are of the relaxed amorphous kind as discussed in Section III. Because of their disorderly
structure it is impossible to identify sharp and precise features in their electronic properties. Nevertheless, we do find that generically these support narrow bands
at the grain boundary both close and away from the Fermi energy, of the kind seen in the tunneling experiments. A next question is what happens when 
the interaction between electrons is switched on. Due to the \ldos enhancement, we expect magnetic moments localized along the grain boundary, to be compared to the results of AFM scans of graphite in Ref.~\onlinecite{Cervenka:2009p2555}. This might provide a concrete model for (existence of) ferromagnetism found in defects on the graphite surface.

We also analyzed in detail the electronic signature of ideal grain boundaries formed from arrays of dislocations. Starting from the perspective of continuum 
field theory revolving around the zero modes associated to Dirac fermions subjected to topological defects, we identified a potentiality of very elegant physics 
associated with grain boundaries. Combining dislocations in a grain boundary, we obtain the striking result that there are localized zero modes decaying as a power-law from the defect, and contributing to the observable \ldos. As we demonstrated, 
the relevancy of these field theoretical results are critically dependent on the details of the microscopic structure of the dislocation core. Murphy's law gets in
the way with the most elementary and natural pentagon-hexagon dislocation core, possibly because this disrupts the topology underneath the continuum limit by 
spoiling the connectivity of the sublattices. However, the ``OCT'' dislocation core appears to be compatible with the continuum theory, and we do find a zero mode structure and the correct power law behavior of the \ldos.
 
We hope that our results will stimulate further experimental research. The challenge appears to find out how to control with great precision the microscopic structure of grain boundaries in graphene in the laboratory. 
 In particular, it would be wonderful when it turns out to be possible to engineer grain boundaries formed from OCT dislocations,  since this would 
form an opportunity to get a closer look at the profound beauty of the zero modes of Dirac fermions.

\begin{acknowledgments}
We want to thank Jiri Cervenka for many stimulating and useful discussions. This work was supported by the Nederlandse Organisatie voor Wetenschappelijk Onderzoek~(NWO).
\end{acknowledgments}

\bibliography{zeromodes}

\end{document}